\documentclass[12pt]{iopart}

\usepackage{amssymb}
\usepackage{color}
\usepackage{graphicx}
\usepackage{cite}

\begin{document}

\title[Critical properties of Bose-Hubbard models]
      {Quantum critical properties of Bose-Hubbard models}

\author{S\"oren Sanders and Martin Holthaus}

\address{Institut f\"ur Physik, Carl von Ossietzky Universit\"at, 
	 D-26111 Oldenburg, Germany}

\ead{martin.holthaus@uol.de}

\date{February 19, 2019}

\begin{abstract}
The Mott insulator-to-superfluid transition exhibited by the Bose-Hubbard 
model on a two-dimensional square lattice occurs for any value of the chemical 
potential, but becomes critical at the tips of the so-called Mott lobes only. 
Employing a numerical approach based on a combination of high-order 
perturbation theory and hypergeometric analytic continuation we investigate
how quantum critical properties manifest themselves in computational practice. 
We consider two-dimensional triangular lattices and three-dimensional cubic 
lattices for comparison, providing accurate parametrizations of the phase
boundaries at the tips of the respective first lobes. In particular, we lend 
strong support to a recently suggested inequality which bounds the divergence 
exponent of the one-particle correlation function in terms of that of the 
two-particle correlation function, and which sharpens to an equality if and 
only if a system becomes critical.

\end{abstract} 


\maketitle 


\section{Introduction}
\label{sec:1}

The application of the theory of critical phenomena~\cite{PelissettoVicari02,
ZinnJustin02,AmitMartinMayor05} to the description of the lambda transition
undergone by liquid helium at about 2.18~K is still confronting physicists with
a seemingly minor, yet annoying challenge: Owing to measurements performed 
under zero-gravity conditions in Earth orbit in order to reduce the rounding 
of the transition caused by gravitationally induced pressure gradients, the 
critical exponent~$\alpha$ describing the singularity of the specific heat 
could be determined with exceptional accuracy~\cite{LipaEtAl03}, finally 
resulting in the value $\alpha_{\rm ex} = -0.0127 \pm 0.0003$. On the other 
hand, theoretical calculations so far have not managed to reproduce this figure
to an accuracy compatible with the experimental uncertainty, despite substantial
effort~\cite{Kleinert00,KleinertFrohlinde01,BurovskiEtAl06,CampostriniEtAl06,
SokolovNikitina16}. To our knowledge, the currently most accurate theoretical 
value has been reported by Campostrini {\em et al.\/}~\cite{CampostriniEtAl06},
based on finite-size scaling analyses of high-statistics Monte Carlo 
simulations and resummations of 22nd-order high-temperature expansions for 
the $\phi^4$ lattice model and the dynamically diluted $XY$ model, giving 
$\alpha_{\rm th} = -0.0151 \pm 0.0003$. Since this comparison of measurement
with calculation constitutes a core test of the renormalization group theory, 
the remaining discrepancy needs to be taken seriously.
 
It is therefore of interest to observe that the quantum phase transition 
exhibited at zero temperature by the Bose-Hubbard model~\cite{FisherEtAl89} 
may contribute to solving this long-standing puzzle. In a grand-canonical 
setting this model depends on two dimensionless parameters, the scaled 
chemical potential $\mu/U$ and the scaled strength $J/U$ of hopping between 
adjacent lattice sites. Keeping $\mu/U$ fixed at a noninteger positive value, 
and increasing $J/U$ from zero to positive values, the $d$-dimensional model 
features a transition from an incompressible Mott insulator to a superfluid
at a certain $(J/U)_{\rm c}$. For almost all $\mu/U$ that transition is mean 
field-like, but at particular multicritical points corresponding to the tips 
of the so-called Mott lobes the transition falls into the universality class 
of the $(d+1)$-dimensional $XY$ model~\cite{FisherEtAl89}. In particular, the 
comparatively simple $2$-dimensional Bose-Hubbard model then belongs to the 
$3$-dimensional $XY$ universality class, the same class which also covers the 
lambda transition of liquid helium. Thus, in principle it should be possible
to deduce the critical exponents of the lambda transition from a 2-dimensional
Bose-Hubbard system~\cite{RanconDupuis11}. 
     
We have previously outlined a strategy to exploit this connection: Utilizing 
numerically executed strong-coupling perturbation theory to high orders, in 
the guise of Eckardt's process-chain approach~\cite{Eckardt09}, one obtains 
divergent-series representations of the model's $k$-particle correlation
functions $c_{2k}(\mu/U,J/U)$; these divergent series can then be analytically
continued with the help of generalized hypergeometric functions, taking up 
a recent suggestion by Mera {\em et al.\/}~\cite{MeraEtAl15} The locus in 
the $J/U$ -- $\mu/U$ parameter plane {\em where\/} the correlation functions 
diverge then marks the phase boundary; the manner {\em how\/} they diverge 
at the multicritical points provides information on the critical 
exponents~\cite{SandersHolthaus17a,SandersHolthaus17b}. Applied to the 
2-dimensional Bose-Hubbard model on a square lattice, the first hard test 
of this strategy indeed has yielded a fairly good estimate of the critical 
exponent~$\beta$~\cite{SandersHolthaus17a}.
Thus, the combination of high-order perturbation theory with hypergeometric
analytic continuation provides an alternative, self-contained ``bootstrap 
technique'' for computing critical exponents that is independent from other
approaches; it is hoped that further refinements of this technique will 
contribute to resolving the current discrepancy between the experimental
and theoretical benchmark data~\cite{LipaEtAl03,CampostriniEtAl06}. 
Here we report further results of our approach not only for the 2-dimensional 
square lattice, but also for the triangular lattice~\cite{TeichmannEtAl10}. 
As an essential consistency check we also consider the Bose-Hubbard model on 
a 3-dimensional cubic lattice, which falls into the 4-dimensional $XY$ class 
and therefore should not become critical at all.

\section{Technical background}
\label{sec:2}

We start by summarizing the technical details of our procedure: Employing the 
repulsion energy~$U$ of a pair of bosons occupying a common lattice site as 
reference energy, we write the Bose-Hubbard Hamiltonian in the dimensionless 
form
\begin{equation}
 	\widehat{H} = 
	\frac{1}{2} \sum_{i} \widehat{n}_i \big(\widehat{n}_i - 1 \big)
      - \mu/U \sum_{i} \widehat{n}_i
      - J/U \sum_{\langle i,j \rangle} \, 
	\widehat{b}_i^{\dagger} \widehat{b}_j^{\phantom \dagger}   
      + \sum_i \eta \big( \widehat{b}_i^{\dagger}
          + \widehat{b}_i^{\phantom \dagger} \big) \; ,   	
\label{eq:HAM}
\end{equation}
where the indices $i$, $j$ label individual sites, and the operators
$\widehat{b}_i^{\phantom \dagger}$ and $\widehat{b}_j^{\dagger}$ are bosonic
annihilation and creation operators, obeying the canonical commutation relation
\begin{equation}
 	[\widehat{b}_i^{\phantom \dagger}, \widehat{b}_j^{\dagger}] = 
	\delta_{ij} \; . 
\end{equation}
Moreover,
\begin{equation}
 	\widehat{n}_i = 
	\widehat{b}_i^{\dagger}\widehat{b}_i^{\phantom \dagger}
\end{equation}
denotes the number operator which counts the particles on site~$i$. Thus, 
the first two terms of the Hamiltonian~(\ref{eq:HAM}) specify, respectively, 
the total scaled on-site repulsion energy, and the system's coupling to the 
chemical potential. The sum in the third term ranges over all pairs of sites
$i$, $j$ that are connected by a tunneling contact, as indicated by the symbol 
$\langle i, j \rangle$, describing the hopping of Bosons from sites~$j$ 
to~$i$. Taken together, these three terms constitute the conventional 
Bose-Hubbard model that has been studied in great detail by many authors,
using a wide variety of techniques~\cite{FreericksMonien96,vanOostenEtAl01,
PolletEtAl04,CapogrossoSansoneEtAl07,PolakKopec07,CapogrossoSansoneEtAl08}.
The fourth term represents spatially homogeneous sources and drains which 
explicitly break particle number conservation; this allows one to introduce
an effective potential depending on the source strength~$\eta$ by means of 
a Legendre transformation~\cite{SantosPelster09}. We expand the intensive
ground-state energy~${\mathcal E}$, that is, the ground-state energy per 
lattice site of the extended model~(\ref{eq:HAM}), as a power series in $\eta$,
\begin{equation}
 	{\mathcal E}(\mu/U, J/U, \eta) = e_0(\mu/U , J/U) 
	+ \sum_{k=1}^\infty c_{2k}(\mu/U, J/U) \, \eta^{2k} \; ,   
\end{equation}
and then apply the process-chain approach~\cite{Eckardt09} for estimating
the $k$-particle correlation functions 
\begin{equation}
	c_{2k}(\mu/U, J/U) = \sum_{\nu=0}^{\infty}
	\alpha_{2k}^{(\nu)}(\mu/U) \Big( J/U \Big)^\nu \; .
\label{eq:SER}
\end{equation}		
To this end, we start from the ground state of the site-diagonal first two 
terms of the system~(\ref{eq:HAM}), and include the other two terms as 
perturbations~\cite{TeichmannEtAl09a,TeichmannEtAl09b,HeilvonderLinden12}.
The computation of $c_{2k}$ then prompts one to account for $k$~creation 
and $k$~annihilation processes; invoking perturbation theory to $n$th order,
with $n \ge 2k$, therefore allows one to include $n - 2k$ tunneling processes. 
After evaluating the series~(\ref{eq:SER}) for given $\mu/U$ to an achievable 
order $\nu_{\rm max}$ in the hopping strength $J/U$, we fit this approximant 
by hypergeometric functions $_{q+1}F_{q}$, allowing us to determine both the 
respective transition point $(J/U)_{\rm c}$ and the exponent 
$\epsilon_{2k}(\mu/U)$ which governs the divergence of $c_{2k}$ at that 
point~\cite{SandersHolthaus17b},  
\begin{equation}
 	c_{2k}(\mu/U, J/U) \sim 
	\Big( (J/U)_{\rm c} - J/U \Big)^{-\epsilon_{2k}(\mu/U)} \; .
\label{eq:DDE}
\end{equation}
From general arguments invoking neither the dimension~$d$ nor the geometry
of the underlying lattice we have inferred the inequality
\begin{equation}
	\epsilon_2(\mu/U) \leq \frac{2}{7} \, \epsilon_4(\mu/U)
\label{eq:INE}
\end{equation}	
which is expected to reduce to an exact equality if and only if the system
becomes critical~\cite{SandersHolthaus17a}. Thus, a comparison of the
divergence exponents $\epsilon_2$ and $\epsilon_4$, computed independently
for a given variant of the model, enables one to detect quantum criticality.
One of the main purposes of the present paper is to lend further support to 
this relation~(\ref{eq:INE}), which indicates quantum criticality through 
equality of both sides, by reporting the results of large-scale numerical 
calculations for three variants of the Bose-Hubbard model.  

\section{Accurate parametrization of phase boundaries}
\label{sec:3}

We first consider the phase boundary $(J/U)_{\rm c}$ which, in accordance 
with the asymptotic relation~(\ref{eq:DDE}), can already be deduced from 
$c_2$ alone. This study is motivated by a suggestion due to Freericks 
{\em et al.\/}~\cite{FreericksEtAl09}, claiming that the two values 
$(\mu/U)_\pm$ limiting a Mott lobe with integer filling factor~$g$ can be 
parametrized by the scaling ansatz~\cite{FreericksEtAl09}
\begin{equation}
	(\mu/U)_\pm = g + A(x) \pm B(x)[\mbox{scaling polynomial}]^{\nu} \; ,	
\end{equation} 
where $x = dJ/U$, and $A(x)$, $B(x)$, and the scaling polynomial are certain
polynomials which have been stated for a 2-dimensional square lattice and 
$g = 1$ as Eq.~(104) in Ref.~\cite{FreericksEtAl09}, and $\nu = 0.6717(1)$
is the critical exponent which governs the correlation 
length~\cite{CampostriniEtAl06}. This is an interesting suggestion, implying
that accurate knowledge of the phase boundary suffices to determine~$\nu$.

\begin{figure}[t]
   	\begin{center}
   	\includegraphics[width=0.73\textwidth]{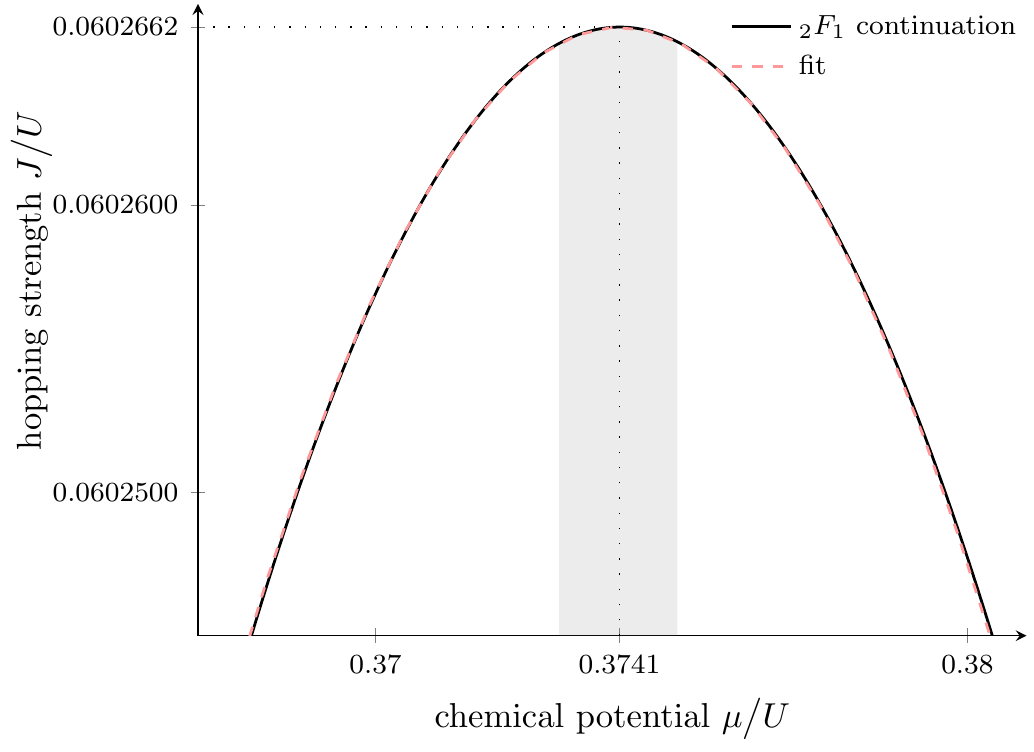}
   	\end{center}
\caption{Tip of the first Mott lobe pertaining to a 2-dimensional square 
	lattice, delineating the boundary $(J/U)_{\rm c}$ between the Mott 
	insulator phase with filling factor $g = 1$ and the superfluid phase. 
	The full line results from hypergeometric continuation, based on 
	$_2F_1$, of the strong-coupling perturbation series evaluated to $9$th 
	order in $J/U$. The dashed line represents the fit~(\ref{eq:F2S}), 
	based on the evaluation of 100 equidistant points within merely the 
	narrow shaded interval $[0.3731,0.3751]$.}      
\label{F_1}
\end{figure}

We therefore have made a precise computation of the phase boundary for the 
first Mott lobe of a 2-dimensional Bose-Hubbard model on a square lattice 
by combining process-chain calculations to $9$th order in $J/U$ with 
hypergeometric analytic continuation based on $_2F_1$~\cite{SandersHolthaus17a,
SandersHolthaus17b}, and have plotted the very tip of this lobe in 
Fig.~\ref{F_1}; here we consider $(J/U)_{\rm c}$ as a function of $\mu/U$. 
We then have taken 100 data points computed in a fairly narrow interval 
of size $\Delta(\mu/U) = 0.002$ around the lobe tip, and have fitted them to 
a function of the form
\begin{equation}
	(J/U)_c = a - b \cdot |\mu/U - c|^y \; ,   
\label{eq:FIT}
\end{equation}
obtaining
\begin{equation}
	(J/U)_{\rm c} = 
	0.0602662 - 0.541909 \cdot |\mu/U - 0.3741|^{2.00018} \; .
\label{eq:F2S}
\end{equation}	
The deviation of the exponent $y = 2.00018$ ``measured'' here from the 
exponent~2 of a perfect parabola is fully accounted for by the limits to our 
numerical accuracy; clearly, the quality of our fit is not compatible with the 
suggested appearance of the critical exponent~$\nu$. Instead, according to our
numerical data the phase boundary $(\mu/U)_\pm$ right at the tip appears to be 
given by a smooth square root function.

\begin{figure}[t]
   	\begin{center}
   	\includegraphics[width=0.73\textwidth]{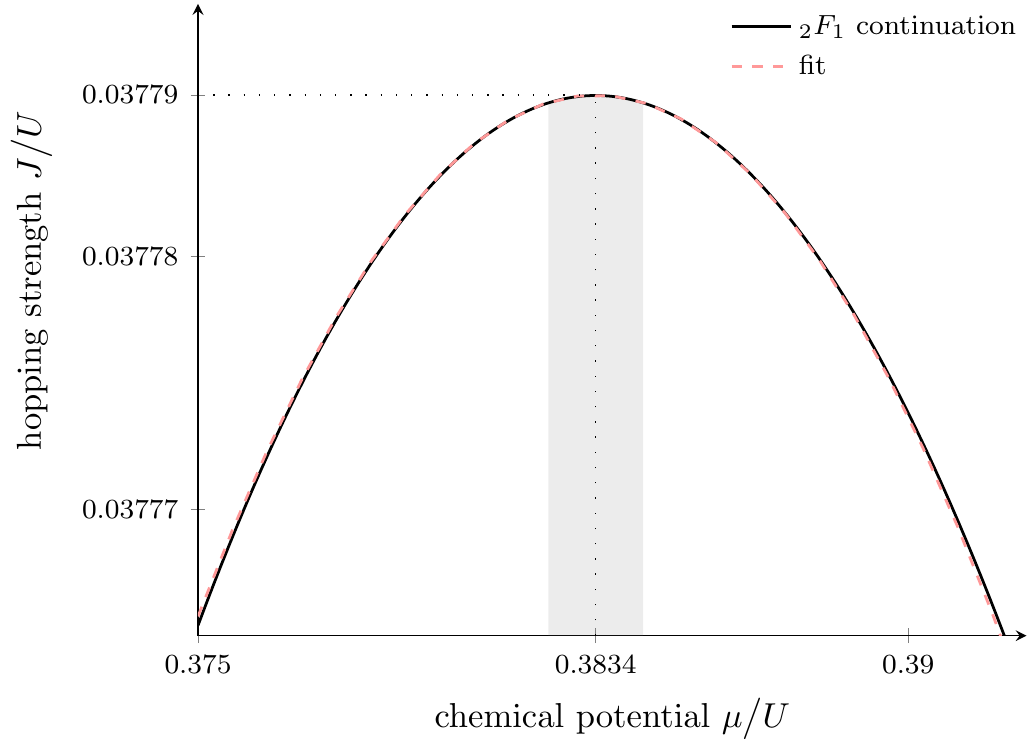}
   	\end{center}
\caption{As Fig.~\ref{F_1}, but for a triangular lattice.
	The full line results from hypergeometric continuation, based on 
	$_2F_1$, of the strong-coupling perturbation series evaluated to $7$th 
	order in $J/U$. The dashed line represents the fit~(\ref{eq:F2T}), 
	based on the evaluation of 100 equidistant points within the shaded 
	interval $[0.3824,0.3844]$.}      
\label{F_2}
\end{figure}

We have also performed such calculations for a triangular lattice: While 
each site of a square lattice is linked by a tunneling contact to four 
nearest neighbors, each site is surrounded by even six nearest neighbors 
on a triangular one. The results are shown in Fig.~\ref{F_2}. As expected 
on the grounds of the respective coordination number, in comparison with 
the square lattice the lobe tip shifts to lower $J/U$ in the triangular 
case~\cite{TeichmannEtAl10}. Again selecting an interval of width 
$\Delta (\mu/U) = 0.002$ around the lobe tip, and taking 100 equidistant 
fit points, the ansatz~(\ref{eq:FIT}) yields  
\begin{equation}
	(J/U)_{\rm c} = 
	0.03779 - 0.2926 \cdot |\mu/U - 0.3834|^{2.00003} \; ,
\label{eq:F2T}
\end{equation}
likewise suggesting that the lobe tip is well approximated by an exact 
parabola, without correction involving a critical exponent.

\begin{figure}[b]
   	\begin{center}
   	\includegraphics[width=0.9\textwidth]{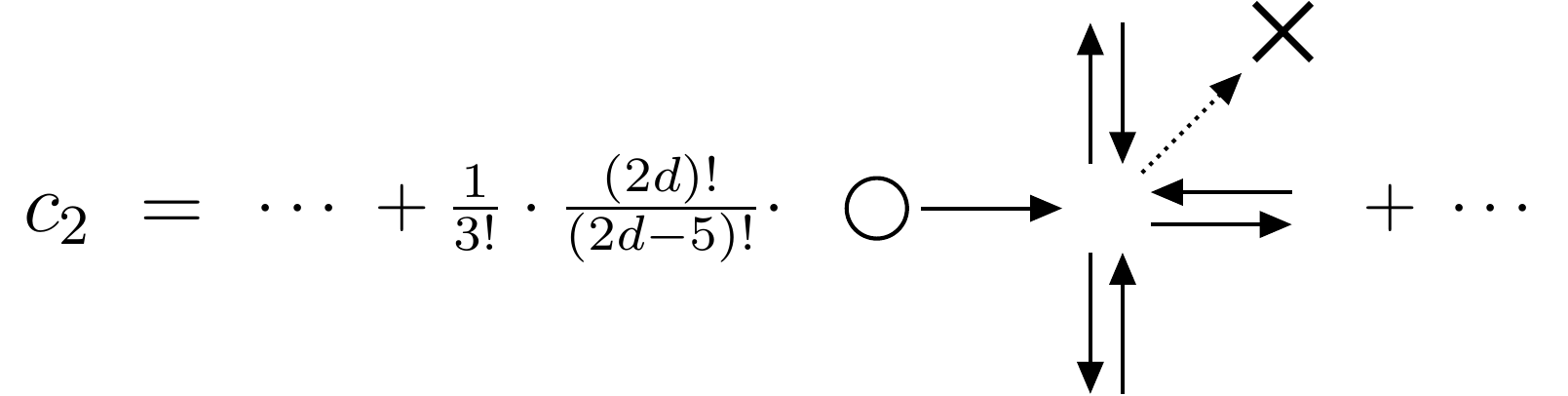}
   	\end{center}
\caption{Diagram contributing to the 8th order in $J/U$ of the perturbation 
	series for the one-particle correlation function $c_2$ of the 
	$d$-dimensional Bose-Hubbard model on a (hyper-)cubic lattice with 
	$d \ge 3$. Arrows denote hopping processes between neighboring lattice 
	sites, the circle symbolizes a creation process, and the cross an 
	annihilation process. The dotted arrow indicates an out-of-plane 
	hopping process which distinguishes the 3-dimensional model from 
	the 2-dimensional one on a square lattice.}          
\label{F_3}
\end{figure}

\begin{figure}[t]
   	\begin{center}
   	\includegraphics[width=0.73\textwidth]{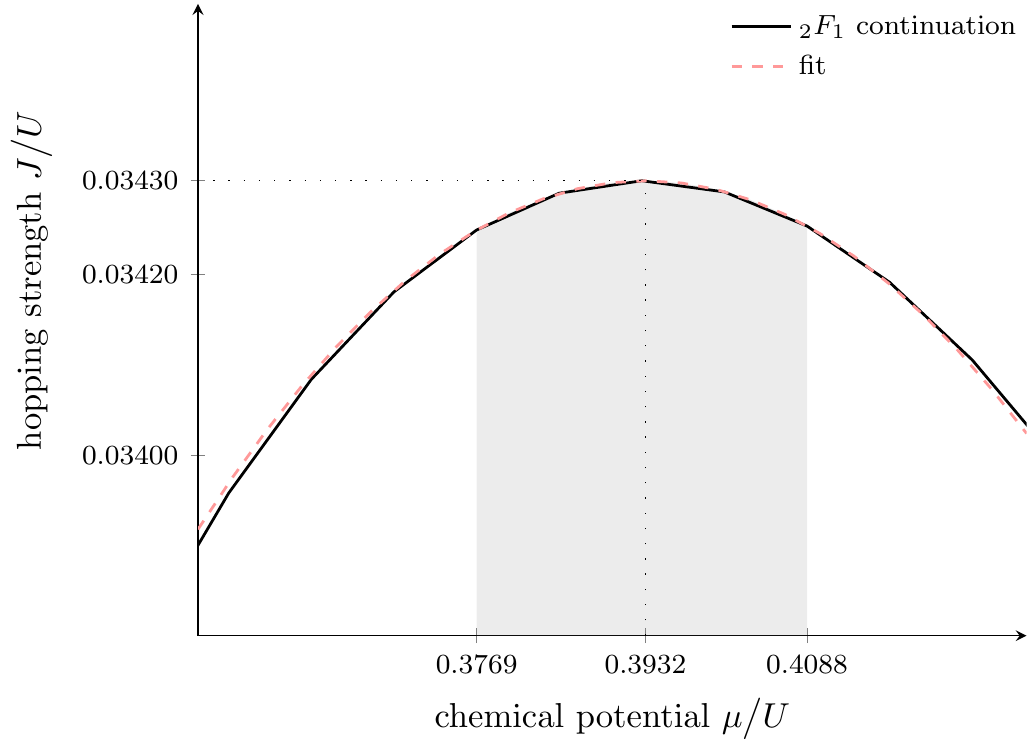}
   	\end{center}
\caption{As Fig.~\ref{F_1}, but for a 3-dimensional cubic lattice.
	The full line results from hypergeometric continuation, based on 
	$_2F_1$, of the strong-coupling perturbation series evaluated to $9$th 
	order in $J/U$. The dashed line represents the fit~(\ref{eq:F3C}), 
	based on the evaluation of merely 5 equidistant points within the 
	shaded interval $[0.3769,0.4088]$. Observe how the scales of the axes
	differ from those in Fig.~\ref{F_1} and Fig.~\ref{F_2}.}      
\label{F_4}
\end{figure}

For comparison, we also have considered a 3-dimensional cubic lattice, for 
which the Mott transition does not become critical~\cite{FisherEtAl89}. From
the perspective of perturbation theory, the process chains defining $c_2$ for 
the 2-dimensional square lattice and those for the cubic lattice are virtually 
the same up to and including the 7th order in $J/U$, with only different weight
factors accounting for the different dimensions. Starting with the order 
$\nu = 8$ additional diagrams begin to appear in three dimensions, such as 
depicted exemplarily in Fig.~\ref{F_3}. The tip of the first Mott lobe is 
displayed in Fig.~\ref{F_4}, together with the fit   
\begin{equation}
	(J/U)_{\rm c} = 
	0.0343041 - 0.210650 \cdot |\mu/U - 0.3932|^{2.00559}	
\label{eq:F3C}
\end{equation}
to the ansatz~(\ref{eq:FIT}). We emphasize that this fit results from the 
evaluation of merely 5 data points taken from a comparatively wide interval 
$\Delta (\mu/U) = 0.0319$, more than 15 times as large as employed in 
Fig.~\ref{F_1}. Yet, the deviation of the observed exponent $y = 2.00559$
from the parabola exponent~2 is not significant.

\section{Manifestation of quantum criticality}
\label{sec:4}

\begin{figure}[t]
   	\begin{center}
   	\includegraphics[width=0.66\textwidth]{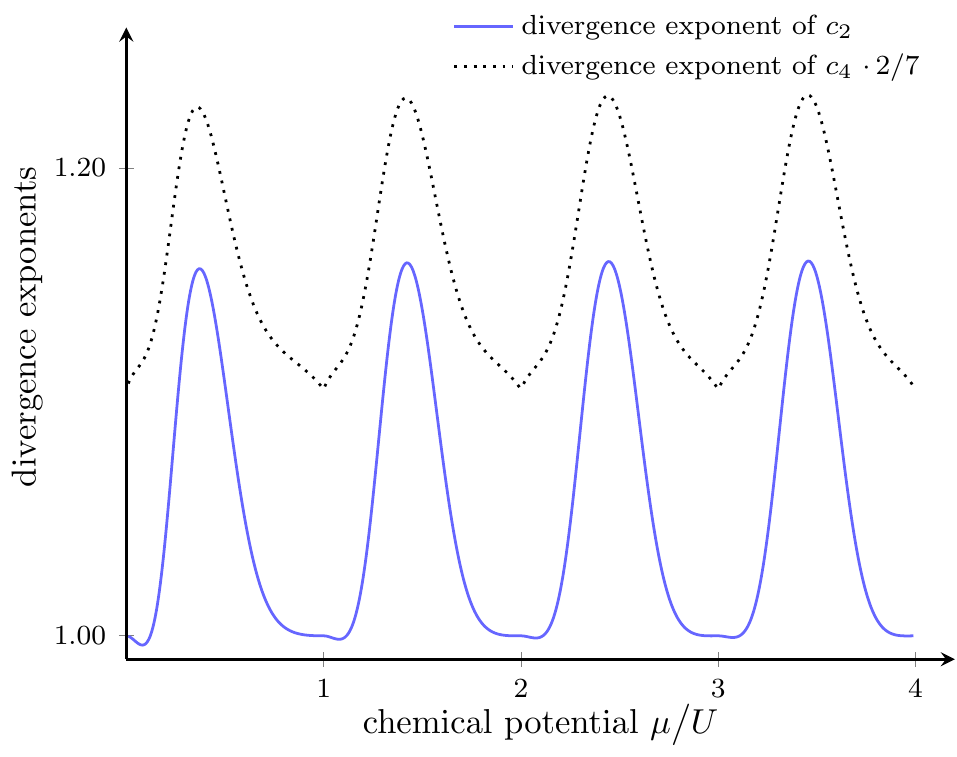}
   	\end{center}
\caption{Comparison of the divergence exponent $\epsilon_2$ of the one-particle
 	correlation function $c_2$ (full line) to $2/7$ times the divergence 
	exponent $\epsilon_4$ of the two-particle correlation function $c_4$
	(dotted), for the 3-dimensional Bose-Hubbard model on a cubic lattice. 
	Here $c_2$ has been evaluated to 9th order in $J/U$, and $c_4$ to 7th 
	order. The inequality~(\ref{eq:INE}) is well satisfied for all values 
	of the scaled chemical potential $\mu/U$, confirming that this model 
	does not become critical.}       
\label{F_5}
\end{figure}

Since it does not appear to be feasible to recognize criticality by 
inspection of the phase boundary alone, we now turn to the more demanding 
criterion~(\ref{eq:INE}), prompting us to evaluate $c_4$ besides $c_2$. We 
start with the 3-dimensional cubic lattice: Figure~\ref{F_5} shows both the 
divergence exponent~$\epsilon_2$, and the divergence exponent~$\epsilon_4$ 
multiplied by $2/7$, for chemical potentials covering the lowest four Mott 
lobes, $0 \leq \mu/U \leq 4$. To produce this figure, the one-particle 
correlation function~$c_2$ has been computed to 9th order in $J/U$, and 
the two-particle correlation function~$c_4$ to 7th order, amounting to the 
evaluation of perturbation theory to 11th order in both cases. Evidently the 
inequality~(\ref{eq:INE}) is well satisfied, equality being excluded for
all $\mu/U$. This is a fairly important consistency check, vindicating the
recognition that the 3-dimensional Bose-Hubbard model does not become critical 
even at the tips of its Mott lobes~\cite{FisherEtAl89,RanconDupuis11}.

\begin{figure}[h]
   	\begin{center}
   	\includegraphics[width=0.66\textwidth]{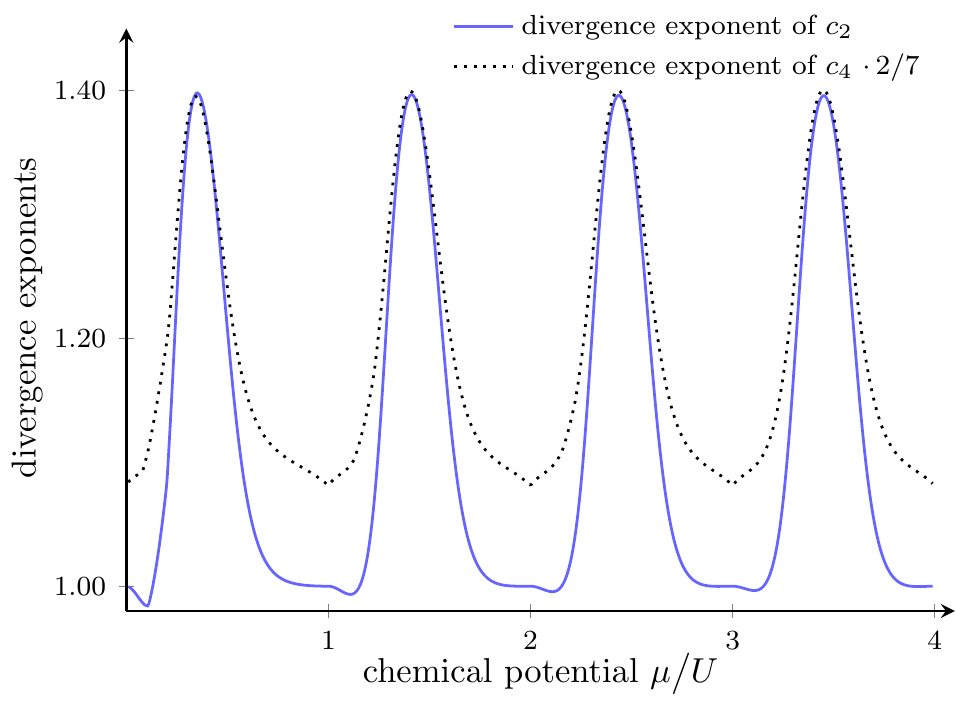}
   	\end{center}
\caption{As Fig.~\ref{F_5}, but for the 2-dimensional Bose-Hubbard model 
	on a square lattice. Here $c_2$ has been evaluated to 10th order 
	in $J/U$, and $c_4$ to 7th order. To numerical accuracy 
	the inequality~(\ref{eq:INE}) actually becomes an equality at the 
	tips of the four Mott lobes considered here, indicating quantum 
	criticality at these points.}       
\label{F_6}
\end{figure}

In contrast, the 2-dimensional models {\em do\/} become critical at the lobe 
tips. The corresponding comparison of $c_2$ and $2/7 \cdot c_4$ for the square 
lattice is shown in Fig.~\ref{F_6}. In marked contrast to Fig.~\ref{F_5}, 
here the two lines touch each other at the lobe tips to within numerical 
accuracy~\cite{SandersHolthaus17a}, so that the inequality~(\ref{eq:INE})
reduces to an equality at these points, thereby heralding criticality.

\begin{figure}[t]
   	\begin{center}
   	\includegraphics[width=0.73\textwidth]{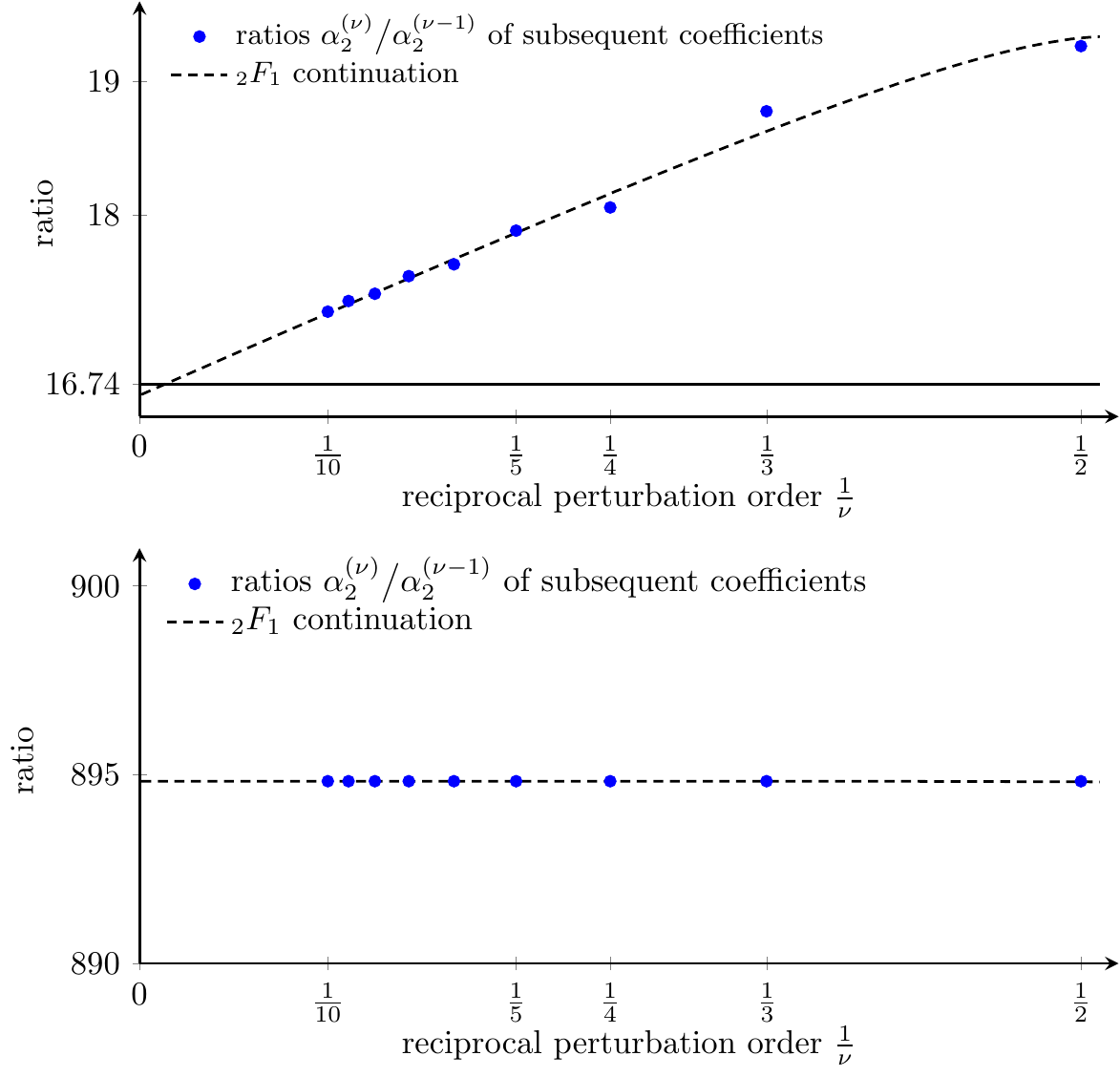}
   	\end{center}
\caption{Ratio $\alpha_2^{(\nu)}/\alpha_2^{(\nu-1)}$ of subsequent coefficients 
	constituting the series representation~(\ref{eq:SER}) for the
	one-particle correlation function $c_2$ of the 2-dimensional 
	Bose-Hubbard model on a square lattice vs.\ $1/\nu$ (dots); together 
	with the corresponding fits to Gaussian hypergeometric functions 
	$_2F_1$ (dashed). The upper panel refers to $\mu/U = 0.3769$, close to 
	the tip of the first Mott lobe; here the flexibility offered by $_2F_1$
	is essential to correctly estimate the limit $1/\nu \to 0$. The lower 
	panel applies to $\mu/U = 0.991$; here the series is very well 
	approximated by a geometric one.}  
\label{F_7}
\end{figure}

For gaining insight into the working principles of hypergeometric analytic 
continuation it is instructive to inspect the behavior of the coefficients 
$\alpha_{2k}^{(\nu)}(\mu/U)$ constituting the series~(\ref{eq:SER}). To this 
end, we plot in Fig.~\ref{F_7} the ratios $\alpha_2^{(\nu)}/\alpha_2^{(\nu-1)}$
for $\mu/U = 0.3769$ (upper panel) and $\mu/U = 0.991$ (lower panel). In the 
latter case, for $\mu/U$ close to the upper edge of the lowest lobe, the 
series representation~(\ref{eq:SER}) of $c_2$ is close to geometric, allowing 
for a comparatively simple and reliable determination of the limit 
$\alpha_2^{(\nu)}/\alpha_2^{(\nu-1)} \to 
1/(J/U)_{\rm c}$~\cite{TeichmannEtAl09a,TeichmannEtAl09b}. In the former case
however, close to the critical lobe tip, the ratios decrease markedly with 
increasing hopping order~$\nu$, and appear to oscillate slightly around a 
common mean trend, so that their fit to a Gaussian hypergeometric function 
$_2F_1$, or to generalized hypergeometric functions $_{q+1}F_q$, indeed is 
indispensable for correctly estimating the limit 
$\nu \to \infty$~\cite{SandersHolthaus17b}.

\begin{figure}[t]
   	\begin{center}
   	\includegraphics[width=0.66\textwidth]{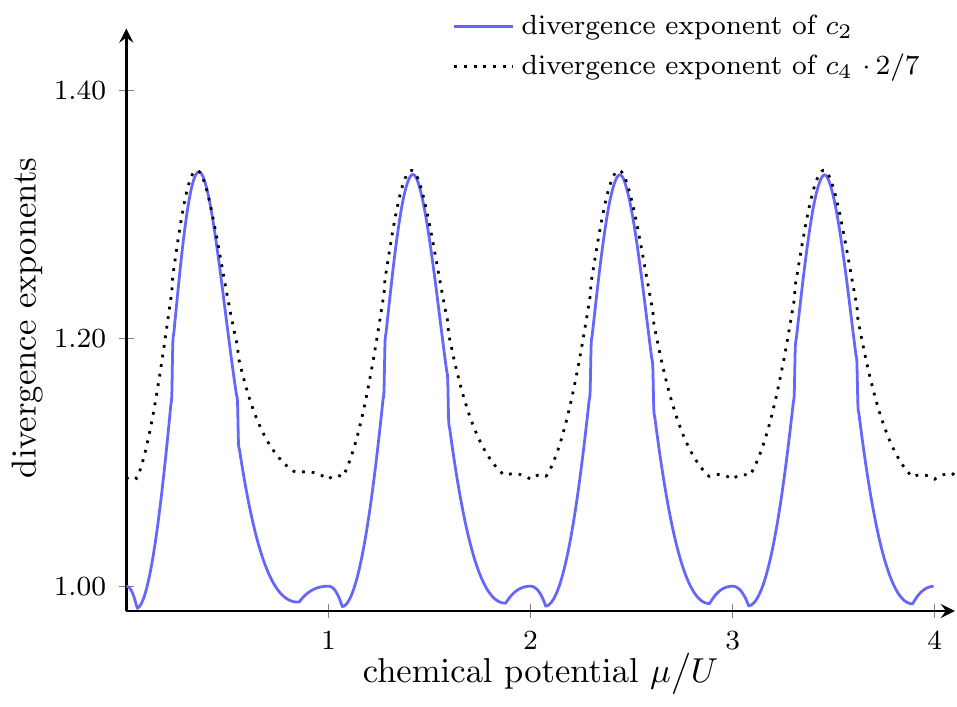}
   	\end{center}
\caption{As Fig.~\ref{F_5}, but for the 2-dimensional Bose-Hubbard model 
	on a triangular lattice. Here $c_2$ has been evaluated to 9th order 
	in $J/U$, and $c_4$ to 6th order. To numerical accuracy 
	the inequality~(\ref{eq:INE}) actually becomes an equality at the 
	tips of the four Mott lobes considered here, indicating quantum 
	criticality at these points.}       
\label{F_8}
\end{figure}

It is now of major interest to check the crucial inequality~(\ref{eq:INE}) 
for the triangular lattice. The numerical data shown in Fig.~\ref{F_8}, 
where once again we have plotted both sides of this inequality, are fairly
convincing indeed: To within numerical accuracy, the inequality~(\ref{eq:INE}) 
is confirmed, with exact equality being reached, again to within numerical 
accuracy, at the lobe tips. Thus, an approach based on high-order perturbation 
theory combined with hypergeometric analytic continuation definitively is 
sensitive to quantum criticality.

\section{Discussion}
\label{sec:5}

The most important findings of the present numerical study are encoded 
in Figs.~\ref{F_5}, \ref{F_6}, and \ref{F_8}, representing a  
computationally rather demanding verification of the divergence-exponent 
inequality~(\ref{eq:INE}) previously surmised in Ref.~\cite{SandersHolthaus17a}:
In general, the divergence exponent~$\epsilon_2$ which characterizes the 
one-particle correlation function~$c_2$ at the phase transition is smaller 
than $2/7$ times the divergence exponent~$\epsilon_4$ of $c_4$, but both sides 
become equal if and only if the system becomes critical.

Yet, there is one more issue that is at stake here. At criticality one 
expects the relation $\beta = (\epsilon_4 - 3\epsilon_2)/2$ for the critical 
exponent $\beta$ of the order parameter~\cite{SandersHolthaus17a}; together
with $\epsilon_4 \stackrel{!}{=} 7\epsilon_2/2$, this yields the remarkable 
identity
\begin{equation}
	\beta = \frac{1}{4} \epsilon_2 \; .
\end{equation}
Indeed, reading off $\epsilon_2$ from the lobe tips seen in Fig.~\ref{F_6} for 
the square lattice with its coordination number $Z = 4$, and dividing by four, 
one obtains fairly good agreement with the expected value $\beta = 0.3486(1)$
reported by Campostrini {\em et al.\/}~\cite{CampostriniEtAl06}.  
While we do no yet reach the numerical accuracy level of this benchmark value, 
this may become feasible after honing our method still further, possibly 
allowing for a rather stringent test of the universality hypothesis of 
statistical physics.  
But the same procedure applied to the data depicted in Fig.~\ref{F_8} for the 
triangular lattice with coordination number $Z = 6$ yields the markedly 
different value
\begin{equation}
	\beta = 0.332(1) \; . 
\end{equation}		  
This is an interesting observation, suggesting that critical exponents of 
the Mott insulator-to-superfluid quantum phase transition in the Bose-Hubbard 
model depend not only on the dimensionality, but also on the coordination
number of the respective lattice. 
The experimental verification of this prediction, possibly with ultracold atoms
in optical lattices~\cite{ZhangEtAl12}, could provide important guidance for
the further development of the theory.  

%
\ack
The numerical calculations underlying this work have been performed on the 
HPC cluster CARL, located at the University of Oldenburg and funded by the 
DFG through its Major Research Instrumentation Programme (INST 184/157-1 FUGG),
and by the Ministry of Science and Culture (MWK) of the Lower Saxony State.
%
%
%
\section*{References}

\end{document}